\documentclass{svproc}
\usepackage{amsmath}
\usepackage{amssymb}
\usepackage{booktabs}
\usepackage[table]{xcolor}
\usepackage{url}

\usepackage{graphicx}
\usepackage{float}
\usepackage{listings}
\usepackage{tabularx}

\begin{document}
\mainmatter              
\title{GenAI-Driven Approach to RISC-V Supply Chain Exploration}
\titlerunning{GenAI-Driven Approach to RISC-V Supply Chain Exploration}  
%

\author{
Nenad Petrovic\inst{1} \and
Andre Schamschurko\inst{1} \and
Yingjie Xu\inst{1} \and
Alois Knoll\inst{1}
}

\authorrunning{N. Petrovic et al.}

\tocauthor{Nenad Petrovic, Andre Schamschurko, Alois Knoll}

\institute{
Chair of Robotics, Artificial Intelligence and Real-Time Systems\\
Technical University of Munich, Munich, Germany\\
\email{\{nenad.petrovic, andre.schamschurko, yingjie.xu, knoll\}@tum.de}
}

\maketitle              

\begin{abstract}
This paper presents an LLM-empowered workflow for RISC-V supply chain analysis, integrating Vision-Language Models (VLMs) and Model-Driven Engineering (MDE) to enable comprehensive, multimodal data-driven insights. The proposed approach addresses the challenges of heterogeneous and unstructured supply chain data by leveraging LLMs for textual understanding and VLMs for extracting information from visual artifacts such as diagrams, tables, and scanned documents. These models collaboratively identify key entities and relationships, which are then organized into a knowledge graph representing supply chain components and their interdependencies. For analytical reasoning, the workflow incorporates MDE techniques and constraint-based modeling to enable formal validation of dependencies, detection of bottlenecks, and assessment of risks. The synergy between LLM- and VLM-based semantic understanding and MDE-based formal analysis supports both exploratory and systematic evaluation of supply chain resilience. A human-in-the-loop mechanism further enables interactive querying and expert validation. The approach is evaluated in RISC-V ecosystem scenarios, demonstrating its effectiveness in generating actionable insights, enhancing transparency, and supporting decision-making in complex semiconductor supply chains.
\end{abstract}
\section{Introduction}
The rapid digitalization of the automotive industry has transformed vehicles into highly complex, software-defined systems that depend on advanced semiconductor technologies. Contemporary vehicles integrate dozens to hundreds of electronic control units (ECUs), supporting functionalities ranging from infotainment to advanced driver assistance systems (ADAS) and autonomous driving. This growing computational demand has intensified the importance of efficient, scalable, and flexible hardware architectures. In this context, RISC-V has emerged as a disruptive alternative to proprietary instruction set architectures, offering an open, modular, and extensible framework for processor design \cite{navik2025riscv}.

The adoption of RISC-V in the automotive domain is motivated by several factors, including reduced licensing costs, design flexibility, and the ability to customize architectures for specific workloads such as real-time processing and edge AI \cite{riscv2025aiopportunities}\cite{cuomo2023riscv}. Automotive original equipment manufacturers (OEMs), Tier 1 suppliers, semiconductor companies, and software vendors are increasingly exploring RISC-V-based solutions to gain greater control over their hardware-software stacks. However, this shift also introduces significant challenges in understanding and managing the associated supply chains, which are inherently more decentralized and heterogeneous than those built around traditional proprietary ecosystems.

Unlike vertically integrated semiconductor ecosystems, the RISC-V landscape is characterized by a distributed network of stakeholders, including intellectual property (IP) core designers, electronic design automation (EDA) tool providers, foundries, packaging and testing facilities, middleware developers, and system integrators \cite{navik2025riscv}. This fragmentation complicates the task of supply chain analysis, as critical information is dispersed across multiple sources, formats, and organizational boundaries. Furthermore, relationships among stakeholders are often dynamic, involving partnerships, licensing agreements, and collaborative development models that evolve over time.

In the automotive sector, these challenges are further amplified by stringent regulatory and safety requirements \cite{andreasyan2026riscv}. Standards such as functional safety (e.g., ISO 26262) impose strict constraints on traceability, verification, and validation across the entire lifecycle of electronic components. Automotive supply chains also exhibit long product lifecycles, often exceeding a decade, which necessitates sustained visibility into component provenance and supplier dependencies. The integration of RISC-V into such an environment requires robust mechanisms for mapping and monitoring complex interdependencies among actors, processes, and technologies.

A key difficulty in analyzing RISC-V automotive supply chains lies in the nature of the available data. Much of the relevant information is embedded in unstructured or semi-structured formats, including technical documentation, whitepapers, standards specifications, press releases, and industry reports. Additionally, supply chain knowledge is frequently conveyed through visual artifacts such as architecture diagrams, block schematics, and ecosystem maps. Traditional analytical approaches, which rely on structured databases and manual curation, are insufficient to capture the richness and scale of this multi-modal information landscape. 

Recent advances in Generative Artificial Intelligence (GenAI) and large language models (LLMs) provide new opportunities to address these limitations. GenAI systems are capable of extracting semantic meaning from unstructured text, identifying entities and relationships, and synthesizing knowledge across heterogeneous sources. Moreover, the integration of vision-language models enables the interpretation of graphical content, facilitating the extraction of structural information from diagrams and visual representations. These capabilities make GenAI a promising foundation for automated supply chain analysis.

In this paper, we propose a GenAI-driven framework for automated analysis of RISC-V supply chains in the automotive domain. The approach leverages multi-modal data ingestion to process both textual and visual inputs, followed by advanced natural language processing techniques to identify key stakeholders, roles, processes, and interactions. The extracted information is systematically structured into a dynamic knowledge graph, which serves as a formalized representation of entities and their relationships, enabling comprehensive exploration, querying, and analysis of the ecosystem. In addition, we adopt principles from Model-Driven Engineering to define metamodels that capture domain concepts and constraints, ensuring consistency, interoperability, and extensibility of the generated knowledge structures. Beyond structural modeling, the approach incorporates multi-swimlane activity diagrams to represent processes and interactions across different stakeholders, enabling clear visualization of responsibilities, workflows, and inter-organizational dependencies. These behavioral models are automatically generated using the PlantUML notation, facilitating reproducibility and seamless integration into engineering toolchains. Recent results on LLM-driven metamodeling indicate the effectiveness of using PlantUML as intermediate representation for LLM-driven domain-specific concept and process extraction starting from freeform textual inputs \cite{petrovic2025llm}\cite{petrovic2025llmEC}. In our case, the synergy of knowledge graphs and model-driven engineering techniques enables the transformation of unstructured, multi-modal data into semantically rich, machine-interpretable, and visually expressive models. By continuously integrating new data sources, the framework supports up-to-date monitoring of supply chain developments, including emerging actors and evolving partnerships, while enabling model-based reasoning and analysis for decision support.

\section{Background and Related Works}
\subsection{Knowledge Graph Generation Using GenAI Paradigms}
Knowledge graphs (KGs) provide explicit representations of entities, relations, and constraints, enabling structured querying, provenance tracking, and incremental updates, which makes them suitable for RISC-V supply chain analysis. Recent work increasingly treats KG construction as a generative information extraction problem, where large language models (LLMs) are used not only to extract entity-relation triplets from unstructured documents, but also to organize them into more structured, schema-aware, and dynamically expandable graphs.

Recent LLM-based KG construction methods evolve from triplet extraction to schema induction and autonomous graph building. Papaluca et al. show that zero- and few-shot LLMs can already perform effective triplet extraction with dynamically retrieved KB context \cite{papaluca2024zero}. CoDe-KG improves extraction quality by combining coreference resolution with syntactic sentence decomposition \cite{anuyah2025automated}, while SAC-KG extends extraction into a generator–verifier–pruner pipeline for automatic domain KG construction \cite{chen2024sac}.

Beyond structural integrity, the field is moving toward schema evolution to handle domains where ontologies are not predefined.
Frameworks such as EDC (Extract, Define, and Canonicalize) couple open extraction with semantic definition and schema canonicalization \cite{zhang2024extract}, while iText2KG~\cite{lairgi2024itext2kg} proposes a plug-and-play, zero-shot method for incremental KG construction that eliminates extensive post-processing. 
AutoSchemaKG further enables schema-evolving KG construction from web-scale corpora \cite{bai2025autoschemakg}.
In more specialized settings, Tree-KG builds hierarchical graphs and iteratively expands hidden knowledge \cite{niu2025tree}, KGGen generates higher-quality KGs from plain text via entity clustering \cite{mo2025kggen}, and LKD-KGC targets domain repositories by inferring inter-document dependencies and autoregressively generating schemas without predefined structures \cite{sun2025lkd}.

Despite these advances, the majority of these pipelines remain text-centric, relying on narrative descriptions to infer relationships.
This limitation is significant for technical domains like RISC-V, where critical provenance evidence is often embedded in visual artifacts (e.g., block diagrams, BOM tables) rather than plain text.
Therefore, recent work moves beyond text-only KG generation toward graph construction from visually rich documents.
VaLiK uses cascaded VLMs and cross-modal consistency verification to construct annotation-free multimodal knowledge graphs \cite{liu2025aligning}.
Query-Driven Multimodal GraphRAG further suggests a practical mechanism for integrating text, figures, and other visual artifacts into structured reasoning pipelines \cite{bu2025query}.

Together, these works suggest that for domains such as RISC-V supply chains, GenAI-based KG generation is becoming increasingly critical.

\subsection{LLMs and VLMs for Supply Chain Analysis}

Supply chains are naturally modeled as multi-tier networks of suppliers, products, facilities, and dependencies. Table \ref{tab:Supply_Chain_Approach2}  summarizes the LLM/VLM-driven supply chain analysis approaches in this subsection.

In recent LLM-based work, a major focus is on constructing structured supply network representations directly from open, unstructured data. AlMahri et al. use zero-shot LLM-based entity and relation extraction to build supply-chain knowledge graphs from public textual sources, showing that LLM-driven extraction can extend visibility beyond directly disclosed supplier links and support richer upstream surveillance \cite{almahri2026enhancing}. Li et al. combine ontology-driven graph construction, graph retrieval-augmented generation, and LLM-based natural language processing for supplier discovery, demonstrating that LLM+KG pipelines can improve the accessibility of manufacturing suppliers \cite{li2025integrating}. Zheng and Brintrup further show that generative AI can strengthen supply-chain visibility not only through extraction from text, but also through relationship prediction over knowledge graphs, improving the recovery of contextualized supply-network links \cite{zheng2025enhancing}.

A second research direction uses LLMs for graph-based risk monitoring and event-centric analysis. Han et al. encode relevant subgraphs of a supply chain KG into textual prompts and use GPT-class models in a multi-turn QA procedure to identify critical supply paths \cite{han2024critical}. Shahsavari et al. propose a lightweight unsupervised framework for identifying supply chain contributing events from news, using LLM-enabled phrase expansion and relevance labeling to support proactive risk monitoring. \cite{shahsavari2024event}. SHIELD combines LLM-driven schema induction and event extraction with graph-based prediction modules and focus on feedback for EV battery supply chain disruption analysis \cite{cheng2024shield}. AlMahri et al. propose an agentic AI approach for supply chain disruption monitoring that couples event detection from unstructured sources with downstream exposure and disruption analysis \cite{almahri2026automating}.

Beyond risk monitoring, OptiGuide \cite{li2023large} and subsequent work on LLMs for supply chain decisions \cite{simchi2025large} show that LLMs can also serve as natural-language interfaces to optimization and planning systems, enabling data discovery, scenario analysis, and faster interaction with mathematical decision tools.

Recent review work confirms that supply chain intelligence with LLMs is expanding rapidly but remains fragmented and largely text-centric, with multimodal supply chain analysis still comparatively underexplored \cite{song2026large,singh2026generative}. This gap is particularly relevant for domains such as semiconductor and RISC-V supply chains, where key evidence is often distributed across diagrams, tables, and scanned technical documentation rather than plain text alone.

The potential to bridge this gap is evidenced by recent progress in general multimodal document understanding, summarized in Table~\ref{tab:Other_Approaches}.
Rather than treating documents as flat text streams, frameworks like DocLLM explicitly model document layout to enhance semantic understanding \cite{wang2024docllm}.
Furthermore, the introduction of structured attention mechanisms in multimodal LLMs has shown that preserving the spatial and hierarchical structure of a document is essential for complex reasoning over non-linear content \cite{liu2025structured}.

These developments underscore a fundamental mismatch between the current state of supply chain AI, which relies on narrative extraction, and the actual nature of technical supply chain documentation.
By integrating the layout-aware capabilities demonstrated in general multimodal research into the specialized domain of RISC-V analysis, it becomes possible to recover critical provenance data that text-only pipelines inherently overlook.

\begin{table*}[ht]
\caption{Comparison of LLM/VLM-driven Supply Chain Analysis Approaches}
\label{tab:Supply_Chain_Approach2}
\centering
\scriptsize 
\setlength{\tabcolsep}{6pt} 
\begin{tabularx}{\textwidth}{l l X l X} 
\toprule
\textbf{Approach} & \textbf{Modality} & \textbf{Methodology} & \textbf{Domain} & \textbf{Key Output} \\ 
\midrule

\rowcolor[gray]{0.95} \multicolumn{5}{l}{\textit{Knowledge Graph Generation \& Discovery}} \\
LLM-KG \cite{almahri2026enhancing}
& Text
& Zero-shot entity/relation extraction
& General SC
& Multi-tier supplier mapping, upstream visibility \\

OntoRAG-SC \cite{li2025integrating}
& Text
& Ontology-driven RAG pipeline
& Manufacturing
& Supplier discovery, semantic retrieval \\

GenAI-LinkPred \cite{zheng2025enhancing}
& Graph
& KG completion + link prediction
& General SC
& Missing link recovery, dependency inference \\

\midrule
\rowcolor[gray]{0.95} \multicolumn{5}{l}{\textit{Risk Monitoring \& Event Analysis}} \\
KG-QA Risk \cite{han2024critical}
& Text+Graph
& Subgraph-to-text prompting + QA
& General SC
& Critical path ID \\

EventLLM \cite{shahsavari2024event}
& Text
& Phrase expansion + relevance labeling
& General SC
& Early disruption signals \\

SHIELD \cite{cheng2024shield}
& Text
& Schema induction + event graph
& EV Battery
& Disruption propagation analysis \\

AgentSC \cite{almahri2026automating}
& Text
& Agentic event-to-exposure pipeline
& General SC
& Disruption signals + assessment \\

\midrule
\rowcolor[gray]{0.95} \multicolumn{5}{l}{\textit{Decision Support \& Optimization}} \\
OptiGuide \cite{li2023large}
& Text
& NL-to-Optimization translation
& General SC
& Planning, solver outputs \\

LLM-Decision \cite{simchi2025large}
& Text
& Scenario analysis + prompt-based eval
& General SC
& Scenario evaluation, insights for planning \\
\bottomrule
\end{tabularx}
\end{table*}

\begin{table*}[ht]
\caption{Summary of LLM/VLM-driven approaches for general multimodal document understanding}
\label{tab:Other_Approaches}
\centering
\scriptsize 
\setlength{\tabcolsep}{6pt} 
\begin{tabularx}{\textwidth}{l l X l X} 
\toprule
\textbf{Approach} & \textbf{Modality} & \textbf{Methodology} & \textbf{Domain} & \textbf{Key Output} \\ 
\midrule

DocLLM \cite{wang2024docllm} 
& Text+Vision
& Text + layout modeling 
& Multimodal docs 
& Structured multimodal document understanding \\

StructAttn-MLLM \cite{liu2025structured} 
& Text+Vision
& Structured attention for layout reasoning 
& Multimodal docs 
& Improved table/diagram reasoning \\

\bottomrule
\end{tabularx}
\end{table*}

\section{Proposed Methodology}

\subsection{Workflow overview}
\begin{figure*}[htbp]
\centering
\includegraphics[width=\textwidth]{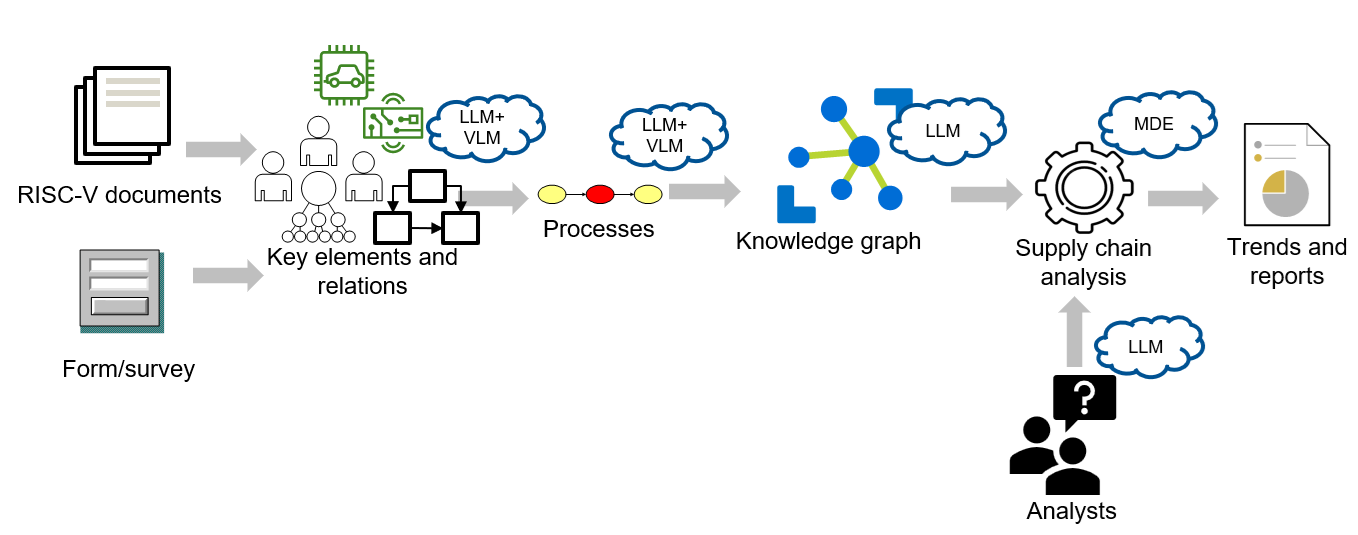}
\caption{Workflow of LLM-empowered event chain-based functional safety by design workflow for automotive.}
\label{fig:sc_analysis}
\end{figure*}

The proposed workflow presents a systematic, AI-driven approach for analyzing the RISC-V supply chain by integrating LLMs, VLMs and MDE. Given the heterogeneous, distributed, and often unstructured nature of supply chain data in semiconductor ecosystems, traditional analysis methods face limitations in scalability and interpretability. This framework addresses these challenges by combining multimodal data understanding (via LLMs and VLMs) with formal modeling and analytical rigor (via MDE). The workflow enables the transformation of diverse data sources into structured knowledge representations, facilitating advanced reasoning, dependency analysis, and decision support. Furthermore, the inclusion of a human-in-the-loop mechanism ensures that domain expertise is incorporated, enhancing the reliability and applicability of the generated insights. The overall flow, described step by step below, illustrates how raw data is progressively converted into actionable intelligence for RISC-V supply chain analysis (depicted in Fig. \ref{fig:sc_analysis}).

\textit{Data acquisition}: Performed by collecting heterogeneous inputs relevant to the RISC-V supply chain, including technical documents, vendor reports, and structured form or survey data. These sources may contain both structured and unstructured information, forming the raw input layer for subsequent processing.

\textit{Key concept identification}: Relevant elements and relationships are extracted from the collected data using a combination of Large Language Models (LLMs) and Vision-Language Models (VLMs). LLMs process textual information to identify entities such as stakeholders, suppliers, components, fabrication facilities, and intellectual property (IP) blocks, while VLMs enable the interpretation of visual content such as diagrams, scanned documents, and tables. This step transforms unstructured inputs into structured representations.

\textit{Process extraction}: The extracted entities and relationships are further analyzed to identify and model supply chain processes. LLMs infer workflows such as design, fabrication, packaging, and distribution, capturing dependencies and interactions among different actors and components within the RISC-V ecosystem.

\textit{Knowledge graph construction}: A knowledge graph is constructed by organizing the identified entities as nodes and their relationships as edges. This graph-based representation provides a unified and machine-interpretable structure of the supply chain, enabling efficient querying and reasoning over complex interdependencies.

\textit{Supply chain analysis}: Supply chain analysis is conducted by integrating Model-Driven Engineering (MDE) with LLM-based reasoning. MDE introduces formal modeling and constraint-based analysis to validate structural dependencies and define market assumptions and conditions in order to perform risk identification, and scenario exploration, as well as detecting bottlenecks, which can help analysts to better understand the market and improve processes accordingly.

\textit{Analysis guidelines}: A human-in-the-loop mechanism is incorporated, allowing analysts to interact with the system using natural language queries. Analysts refine assumptions and guide the analytical process, ensuring domain expertise is effectively integrated with automated intelligence. Freeform text given by analysts is transformed to knowledg graph queries and asserts thanks to LLM.

\textit{Results summarization}: The final stage produces actionable outputs, including trend analyses, risk assessments, and comprehensive reports. These insights support decision-making, enhance transparency, and improve resilience within the RISC-V supply chain.

\subsection{PlantUML Activity Diagram-Driven Process Extraction}
\begin{figure*}[htbp]
\centering
\includegraphics[width=\textwidth]{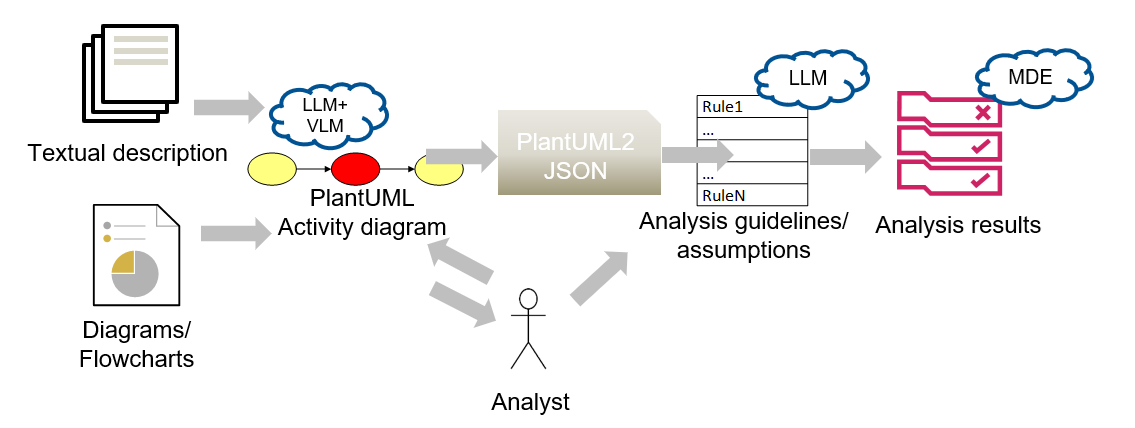}
\caption{Workflow of GenAI-enabled RISC-V process extraction.}
\label{fig:process_analysis}
\end{figure*}

The proposed workflow for Process Extraction using PlantUML notation (depicted in Fig. \ref{fig:process_analysis}, presents a structured pipeline that transforms heterogeneous inputs into formal, machine-readable process models, while emphasizing PlantUML as an intuitive and easily visualizable representation of the extracted workflows. It combines AI-based extraction, structured transformation, and rule-based validation, with PlantUML serving as a key enabler for simple, transparent, and effective visualization of process models.

The process starts with multi-modal input acquisition, where both textual descriptions (such as manufacturign and other process descriptions), as well as visual artifacts such as diagrams or flowcharts are collected. These complementary sources provide a comprehensive view of the process, but require unified interpretation.

Next, GenAI module processes the inputs. The LLM extracts process elements— such as activities, decisions, and control flows—from textual data, while the VLM interprets the  provided diagrams. The outputs are merged into a consistent intermediate representation capturing the overall process semantics. 

The underlying prompt (both system instructions and user-perspective prompt) can be summarized as follows:

\textit{Prompt 1 - PlantUML extraction: (System) "You are generating PlantUml activity diagram about RISC-V supply chain processes without comments and without explanations. Activity diagram should support multiple swimlanes and interaction between actors."
(User) "Update given PlantUML diagram [PlantUML activity diagram text] based on given [textual description]/[input diagram]."}

This representation is summarized as PlantUML activity diagram, which plays a central role in the workflow. PlantUML offers a lightweight textual syntax that can be directly rendered into clear, standardized diagrams, making it particularly suitable for both automated generation and human interpretation. This step ensures that the extracted process is not only formally defined but also immediately visualizable, enabling quick validation and communication among stakeholders.

To support further automation, the PlantUML model is converted into a PlantUML2JSON format, allowing structured access to workflow elements for downstream processing. This transformation bridges the gap between human-readable diagrams and machine-oriented representations.

Subsequently, the model undergoes rule-based analysis using predefined guidelines and assumptions (Rule1 … RuleN). These rules ensure syntactic correctness, semantic consistency, and domain alignment. LLM-assisted reasoning may also be applied to handle ambiguities or incomplete information.
o complement the knowledge graph--based analysis, this work introduces a rule-based framework for examining the ordering and coordination of activities within the RISC-V supply chain. Activities are modeled as ordered events with precedence constraints expressed in the form $A_i$ \textit{before} or \textit{after} $A_j$, enabling the identification of violations in expected execution sequences. To capture conditional execution paths, we extend these constraints with two additional variants: \textit{after-true} and \textit{after-false}. These operators allow the specification of branching behavior, where the execution of an activity depends on the outcome of a prior condition, reflecting decision points such as validation or compliance checks. Furthermore, to support distributed scenarios involving multiple actors, we introduce role-annotated rules of the form $R_k : A_m$, indicating that action $A_m$ is performed by role $R_k$. This enables explicit modeling of coordination across entities such as designers, vendors, and manufacturers. Both free-form, user-defined rules and rules derived from domain catalogs are supported. These are automatically transformed into a formal representation using large language models (LLMs), ensuring consistency and interpretability. Integrated with the knowledge graph, this rule-based layer enables systematic detection of unsafe or inconsistent supply chain management behaviors at design time.

The underlying prompt takes into account both the PlantUML activity diagram of the process and freeform textual rule in the following form: 
\textit{Prompt 2 - Rule generation: For given rules: [textual rules] and activity diagram: [PlantUML activity diagram text], generate list of rules mapped to appropriate ordering constraints: before, after, after-true, after-false. Identitfy activities (A) and roles for included activities (as R:A).
}

The overall workflow incorporates a human analyst in the loop, enabling refinement of extracted models, adjustment of rules, and incorporation of domain expertise. This iterative interaction enhances the accuracy and reliability of the results.

Finally, the system produces analysis results. The strong visual nature of PlantUML makes the results easy to interpret, review, and communicate, while the analysis is performed against the simple JSON form of the diagram enables convenient integration with MDE tools for simulation, verification, and further reasoning.

\subsection{Knowledge Graph-Driven RISC-V Supply Chain-Related Concept Exploration}

\begin{figure*}[htbp]
\centering
\includegraphics[width=\textwidth]{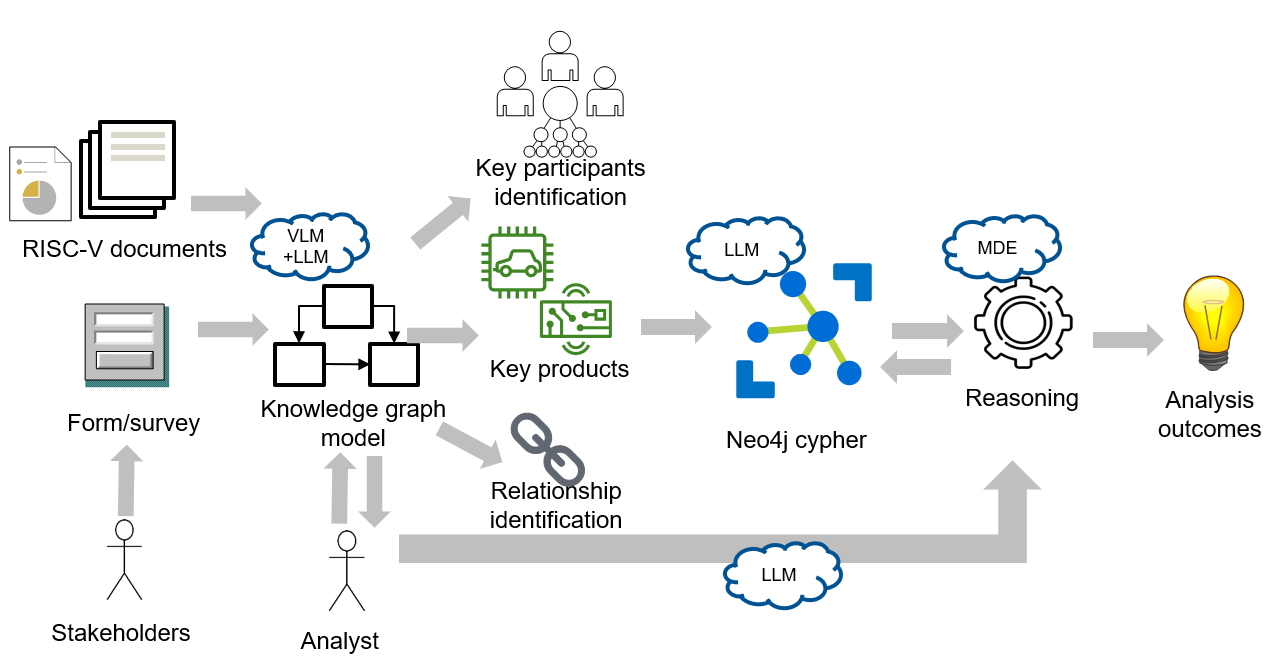}
\caption{Workflow of knowledge graph-driven, GenAI supported RISC-V supply chain concepts exploration.}
\label{fig:kg_analysis}
\end{figure*}

This workflow (depicted in Fig. \ref{fig:kg_analysis}) aims to provide the means for analyzing the RISC-V supply chain through the integration of knowledge graphs, Neo4j graph databases, and large language models (LLMs). The approach is designed to address the inherent complexity and fragmentation of the RISC-V ecosystem by transforming heterogeneous data into a structured, queryable, and reasoning-ready representation.

Inputs from diverse data sources, including unstructured technical documents such as RISC-V specifications, vendor reports, and white papers, as well as structured inputs collected from stakeholders via surveys and forms are provided. These sources collectively capture both technical and organizational aspects of the supply chain. Preprocessing steps, including data normalization, cleaning, and deduplication, are applied to ensure consistency and reliability across inputs.

To extract meaningful knowledge from these data sources, the methodology employs a combination of large language models (LLMs) and vision-language models (VLMs). LLMs are used to process textual information and identify relevant entities such as companies, products, and technologies, along with their attributes and contextual relationships. VLMs complement this process by enabling the interpretation of visual elements, including diagrams and architectural representations commonly found in hardware documentation. The result is a structured set of entities, attributes, and relationships that describe the RISC-V ecosystem.

\textit{Prompt 3 - Knowledge graph extraction: (System) "You are generating Neo4j model about supply chain key elements - stakeholders, products, their relationships and attributes - no comments, delimiters and explanations. Just extract node labels, properties and relationships, without constriants and indices. Example of simple output is given as [example knowledge graph model]."
(User) "Update given knowledge graph model [Neo4j graph model] based on given [textual description]/[input diagram]."}

The extracted knowledge is then organized into a knowledge graph, where entities are represented as nodes and their interactions as edges. The graph schema is designed to capture key supply chain dimensions, including participants (e.g., semiconductor companies and tool providers), products (e.g., processors, IP cores, and software toolchains), and interdependencies (e.g., supply relationships, licensing agreements, and integration pathways). This structured representation enables a holistic view of the ecosystem and supports advanced analytical operations.

Following graph construction, the methodology performs entity and relationship analysis to identify critical components of the supply chain. Key participants are identified based on their roles and connectivity within the network, while key products are determined by their strategic importance and dependency relationships. The classification of relationships—such as supplier-customer links, partnerships, and technological dependencies—further enhances the interpretability of the graph.

The knowledge graph is implemented and managed using Neo4j, a graph database that enables efficient storage, traversal, and querying of complex relationships. Using the Cypher query language, analysts and automated systems can explore multi-hop dependencies, detect potential bottlenecks, and trace supply chain paths across multiple entities. This capability is particularly valuable for identifying critical vulnerabilities and assessing systemic risks.

To enhance analytical capabilities, the framework incorporates a reasoning layer that combines LLM-driven query generation with model-driven engineering (MDE) principles. In this layer, LLMs assist in formulating queries based on analyst's textual guidelines, interpreting graph outputs, and generating insights, while MDE introduces formal rules and constraints that guide consistent and domain-aware reasoning. This hybrid approach enables both exploratory and rule-based analysis of the supply chain. 

For this purpose, we use the following prompt pattern: 
\textit{Prompt 4 - Generate a Cypher query from the [user text] using the provided Neo4j schema [current graph schima]."}

Finally, the system produces actionable analysis outcomes, including identification of critical suppliers, mapping of dependency structures, and detection of potential risks within the RISC-V ecosystem. The methodology supports a human-in-the-loop paradigm, where analysts and stakeholders iteratively refine the data, graph structure, and analytical results. This continuous feedback loop ensures that the model remains accurate, up-to-date, and aligned with real-world dynamics.

Overall, the proposed approach provides a scalable and intelligent framework for supply chain analysis, leveraging the combined strengths of knowledge graphs, graph databases, and artificial intelligence to deliver deep insights into the evolving RISC-V landscape.

\subsection{Implementation Overview}

\begin{figure*}[htbp]
\centering
\includegraphics[width=\textwidth]{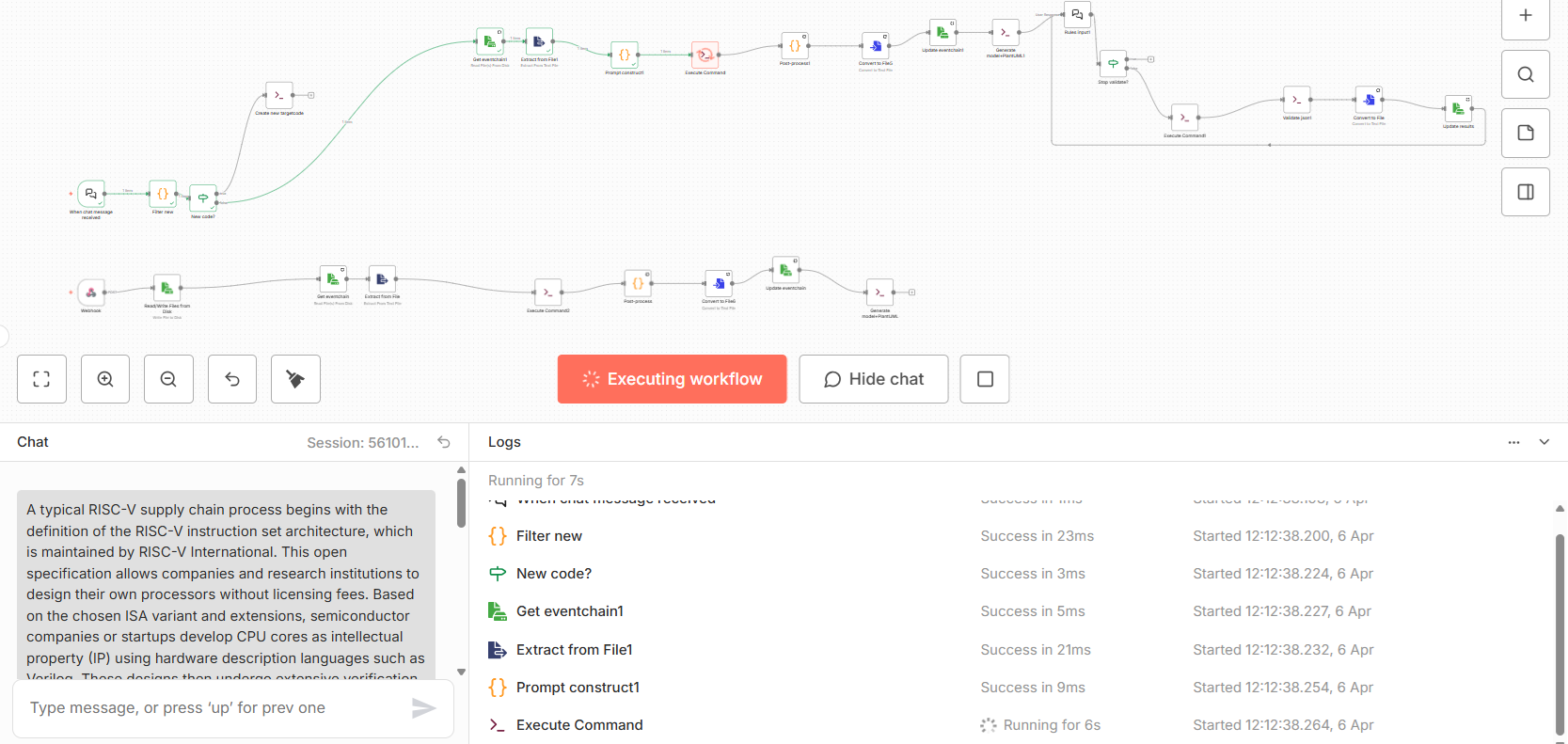}
\caption{Workflow implementation in n8n.}
\label{fig:n8n_impl}
\end{figure*}

When it comes to implementation of the underlying workflows, we rely on openly available n8n tool, which gives us the ability to integrate various processing steps - from input handling to GenAI model prompting and model-driven analysis. Fig \ref{fig:n8n_impl} shows screenshot of the workflow for supply chain process extraction. 

In general, there are two possible ways of user interaction that we leverage in our implementation: 1) interactive chat - workflow execution is triggered after user's textual input; interaction between user and the workflow is based on textual message exchange, where different user responses could potentially lead to alternative execution flows 2) webhook - gives the ability to deploy externally accessible web services invoked via HTML forms, while submission of HTTP request (via POST method) with corresponding input data (such as diagrams or files) acts as workflow execution trigger.

Our n8n-based workflows use command execution nodes in order to parametrize and invoke Python scripts which are responsible for GenAI model prompting, building upon the good practices from our previous work on LLM toolchains \cite{petrovic2025genai}. Models themselves are deployed locally, within our chair's cluster, 
The chair inference server uses a OpenAI API-compatible interface, while Inference server configuration is based on three H200 NVL (141 GB) GPUs. Currently, the following models are supported locally: openai/gpt-oss-120b, Qwen/Qwen3-30B-A3B-Thinking-2507, openai/gpt-oss-120b, mistralai/Magistral-Small-2509 and Qwen/Qwen3.5-122B-A10B.

On the other side, for knowledge graph construction, we make use of Neo4j version 4.1.13.

\section{Experiments and Results}
In what follows, we present an example textual description about RISC-V supply chain that will be used for both the knowledge graph and business process extraction.

Description: \textit{A typical RISC-V supply chain process begins with the definition of the RISC-V instruction set architecture, which is maintained by RISC-V International. This open specification allows companies and research institutions to design their own processors without licensing fees. Based on the chosen ISA variant and extensions, semiconductor companies or startups develop CPU cores as intellectual property (IP) using hardware description languages such as Verilog. These designs then undergo extensive verification and simulation using electronic design automation tools to ensure correctness and compliance with the standard.
Once validated, the CPU core is integrated into a larger system-on-chip (SoC) alongside components like memory controllers, input/output interfaces, and specialized accelerators. The complete chip design is then sent to semiconductor foundries such as TSMC or Samsung Electronics for fabrication, where silicon wafers are manufactured using advanced lithography processes. After fabrication, the chips are packaged, tested, and validated for quality and performance.
Finally, the processors are delivered to original equipment manufacturers (OEMs), who integrate them into end products such as embedded systems, IoT devices, or computing platforms. Alongside the hardware flow, a parallel software ecosystem—including operating systems like Linux, compilers, and development tools—ensures that the hardware can be effectively used in real-world applications. This modular and open approach distinguishes the RISC-V supply chain, enabling multiple independent players to participate at different stages and accelerating innovation across the ecosystem.}

\subsection{Knowledge Graph Extraction}

\begin{figure*}[htbp]
    \centering
    \includegraphics[width=\textwidth]{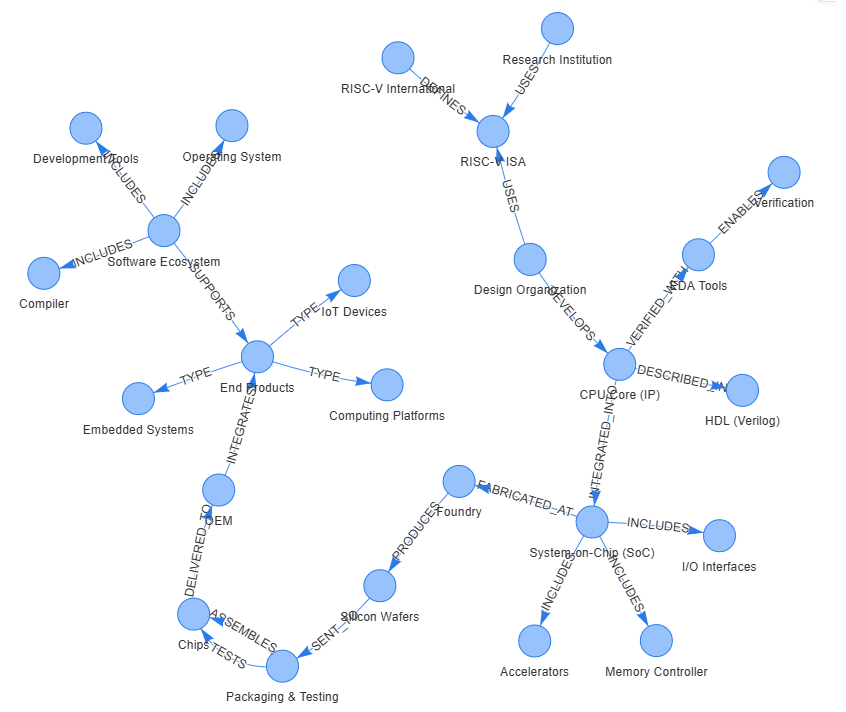}
    \caption{Outcome of knowledge graph extraction in Neo4j style visualization.}
    \label{fig:kg_ex1}
    \end{figure*}

Fig. \ref{fig:kg_ex1} depicts the extracted knowledge graph represented using the Neo4j notation.
   
\subsection{Business Process Analysis}

\begin{figure*}[htbp]
    \centering
    \includegraphics[width=\textwidth]{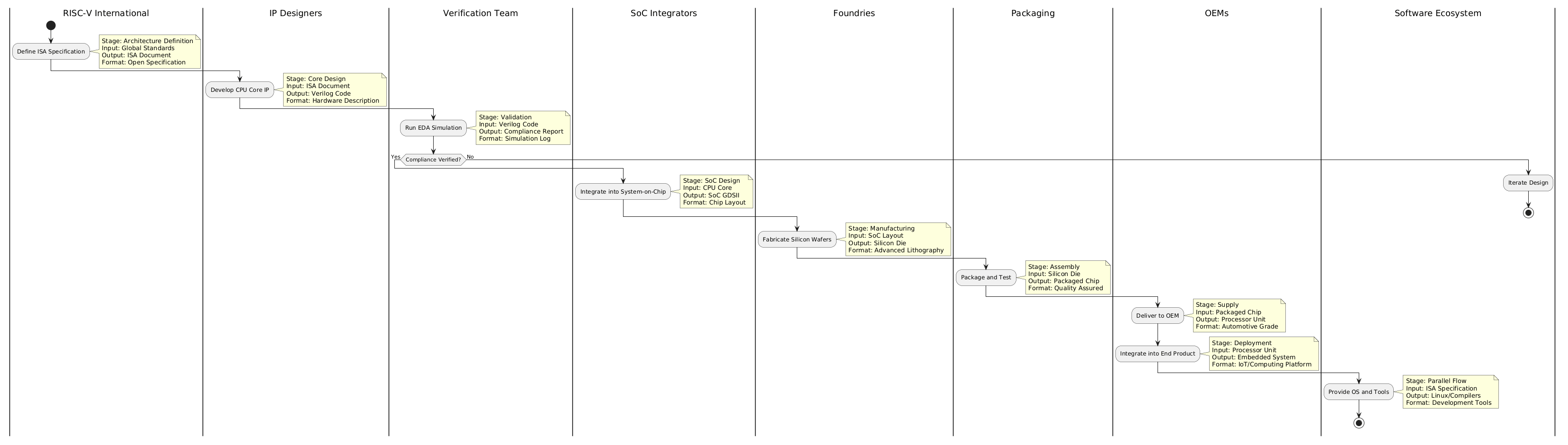}
    \caption{Outcome of text analysis in form of PlantUML diagram.}
    \label{fig:scenarios_proc1}
    \end{figure*}

Fig. \ref{fig:scenarios_proc1} shows the extracted process formalized in PlantUML activity diagram notation.
   
On the other side, Table \ref{tab:validation_rules_param} provides overview of RISC-V supply chain analysis rules which can be leveraged for validation.
\begin{table}[t]
\caption{Validation Rules with Instantiated Activities from the RISC-V Supply Chain}
\label{tab:validation_rules_param}
\centering
\begin{tabular}{p{0.95\columnwidth}}
\hline
\textbf{Rule ID and Description} \\
\hline

\textbf{Rule 1: ISA Definition Precedence} -- Define ISA Specification \textbf{before} Develop CPU Core IP. \\

\textbf{Rule 2: Design-Verification Ordering} -- Develop CPU Core IP \textbf{before} Run EDA Simulation. \\

\textbf{Rule 3: Verification Dependency} -- Run EDA Simulation \textbf{before} Compliance Verification Decision. \\

\textbf{Rule 4: Conditional Integration (after-true)} -- Integrate into System-on-Chip \textbf{after-true} Compliance Verified = Yes. \\

\textbf{Rule 5: Conditional Redesign (after-false)} -- Develop CPU Core IP \textbf{after-false} Compliance Verified = No. \\

\textbf{Rule 6: SoC to Fabrication Constraint} -- Integrate into System-on-Chip \textbf{before} Fabricate Silicon Wafers. \\

\textbf{Rule 7: Fabrication to Packaging Ordering} -- Fabricate Silicon Wafers \textbf{before} Package and Test. \\

\textbf{Rule 8: Testing to Delivery Sequence} -- Package and Test \textbf{before} Deliver to OEM. \\

\textbf{Rule 9: OEM Integration Ordering} -- Deliver to OEM \textbf{before} Integrate into End Product. \\

\textbf{Rule 10: Role-Based Design Assignment} -- $R_{\text{IP Designers}}$: Develop CPU Core IP. \\

\textbf{Rule 11: Role-Based Fabrication Assignment} -- $R_{\text{Foundries}}$: Fabricate Silicon Wafers. \\

\textbf{Rule 12: Role-Based Integration Assignment} -- $R_{\text{OEMs}}$: Integrate into End Product. \\

\textbf{Rule 13: Parallel Software Flow} -- Provide OS and Tools \textbf{parallel to} Hardware Integration and Deployment. \\

\textbf{Rule 14: Safety and Consistency Constraint} -- Package and Test must not occur \textbf{before} Fabrication to avoid invalid production states. \\

\hline
\end{tabular}
\end{table}

\section{Experiments and Results Overview}

The evaluation of locally deployable LLM and VLM for RISC-V supply chain analysis considers both knowledge graph extraction and process analysis from textual and visual inputs. The textual model \texttt{openai/gpt-oss-120b} and the visual model \texttt{Qwen/Qwen3.5-122B-A10B} are evaluated on a representative RISC-V supply chain description and corresponding diagrams. 

A reference knowledge graph is constructed comprising 25 concepts, 23 relationships, and 8 attributes. In parallel, a reference process model is derived from the diagram, consisting of 8 participants, 11 activities, 11 relationships, 9 artifacts, and 5 rules. The ground truths are produced with help of GPT-5.3. These values serve as ground truth for quantitative evaluation.

For knowledge graph extraction, the LLM achieves strong performance across all aspects. Specifically, it correctly extracts 22 out of 25 concepts (88\%), 19 out of 23 relationships (83\%), and 6 out of 8 attributes (75\%). The VLM extracts 18 out of 25 concepts (72\%), 15 out of 23 relationships (65\%), and 5 out of 8 attributes (63\%). The results indicate that LLMs are highly effective in identifying semantically rich entities and explicit dependencies from textual descriptions, while VLMs complement this by capturing structural elements from diagrams. However, attribute extraction remains challenging for both models due to the implicit nature of many properties.

When it comes to Cypher queries generation for purpose of supply chain analysis, an overall end-to-end correct query accuracy of 75\% over an evaluation set of 20 queries means that 15 queries produce fully correct results, while 5 queries exhibit partial or incorrect outputs. A query is considered correct only if it is syntactically valid and returns results fully consistent with the ground truth graph. This indicates that a clear majority of the generated queries are both executable and semantically accurate, with the remaining errors typically caused by missing entities, incomplete relationships, or imperfect attribute extraction that affect the final query results.

For process analysis, the LLM demonstrates strong capability in reconstructing the supply chain workflow, identifying 10 out of 11 activities (91\%), all 8 participants (100\%), 9 out of 11 relationships (82\%), and 7 out of 9 artifacts (78\%). In contrast, the VLM identifies 8 out of 11 activities (73\%), 6 out of 8 participants (75\%) and 7 out of 11 relationships (64\%). The LLM also generates 4 out of 5 rules (80\%), capturing key constraints such as the dependency of fabrication on validated designs and the sequencing of packaging after wafer production. Rule generation is not performed using VLMs. 

When it comes to execution time, average time of LLM response generation was around 11.5s, while it was close to 14s for VLM. All the experiment outcomes are based on average of 5 executions.

\begin{table}[h]
\centering
\caption{Quantitative Results for Knowledge Graph and Process Extraction}
\small
\begin{tabular}{lccc}
\toprule
\textbf{Aspect} & \textbf{Ground Truth} & \textbf{LLM (gpt-oss-120b)} & \textbf{VLM (Qwen3.5)} \\
\midrule

Concepts & 25 & 22 (88\%) & 18 (72\%) \\
Relationships (KG) & 23 & 19 (83\%) & 15 (65\%) \\
Attributes & 8 & 6 (75\%) & 5 (50\%) \\

Participants & 8 & 8 (100\%) & 6 (75\%) \\
Queries & 20 & 15 (75\%) & N/A \\

Activities & 11 & 10 (91\%) & 7 (64\%) \\
Relationships (Process) & 11 & 9 (82\%) & 8 (73\%) \\
Rules & 5 & 4 (80\%) & N/A \\

\bottomrule
\end{tabular}
\end{table}

Overall, the results highlight that LLMs significantly outperform VLMs in semantic extraction and reasoning tasks, particularly for relationships, attributes, and rule generation. VLMs provide complementary strengths in identifying structural components and process elements from visual representations but exhibit limitations in handling abstract concepts, conditional dependencies, and iterative flows. The integration of both modalities enables a more comprehensive analysis of the RISC-V supply chain, with LLMs serving as the primary reasoning component and VLMs enhancing structural understanding from diagrams.

\section{Conclusion}
The combined use of LLMs and VLMs enables a comprehensive approach to supply chain analysis covering knowledge graph and process extraction, as it was shown for RISC-V domain in our papaer. LLMs provide strong semantic understanding and reasoning capabilities, while VLMs enhance the extraction of structural information from visual inputs. This multimodal integration supports robust knowledge graph construction and process modeling. LLM is highly effective in reconstructing supply chain workflows and generating formalizable rules, enabling reasoning over dependencies and constraints. VLMs provide complementary support by identifying structural process elements from visual inputs but are not suitable for rule generation due to limitations in abstract reasoning. On the other side, based on our experiments it can be concluded that locally deployable models in combination with additional steps under several constraints (simpler intermediate representations, iterative approach, model-to-model transformations and result validation) are quite close to start-of-art commercial models. Relying on locally deployable models instead of external services could be highly beneficial in more sensitive RISC-V adoption domains, such as automotive. In future works, it is planned to focus on agentic GenAI approach adoption in order to analyse supply chain from perspective of different stakeholder and participant roles, which would be helpful for deeper understanding of RISC-V market aspects.

\subsubsection{Acknowledgments.} This work has received funding from the European Chips Joint Undertaking under Framework Partnership Agreement No. 101194371 (Rigoletto) including national funding from the German Federal Ministry of Research, Technology and Space (BMFTR). The responsibility for the content of this publication lies with the authors.

\bibliographystyle{splncs03_unsrt}
\bibliography{refs} 

@inproceedings{navik2025riscv,
  title={RISC-V: Redefining the Future of Computing, Architecture, Innovations, and Beyond},
  author={Navik, A. P. and Tiwari, S. K. and Anand, V. and Yadav, B. and Park, J. and Sung, H. J.},
  booktitle={2025 8th International Conference on Electronics, Materials Engineering \& Nano-Technology (IEMENTech)},
  pages={1--5},
  year={2025},
  organization={IEEE},
  address={Kolkata, India},
  doi={10.1109/IEMENTech65115.2025.10959584}
}

@misc{riscv2025aiopportunities,
  title={RISC-V for Automotive AI Use Cases: Opportunities and Challenges},
  author={{RISC-V International}},
  year={2025},
  howpublished={White paper},
  url={https://riscv.org/wp-content/uploads/2025/04/RISC-V_AIOpportunitiesChallenges_042825.pdf},
  note={Accessed: 2026-04-25}
}

@inproceedings{cuomo2023riscv,
  title={Towards a RISC-V Open Platform for Next-generation Automotive ECUs},
  author={Cuomo, L. and others},
  booktitle={2023 12th Mediterranean Conference on Embedded Computing (MECO)},
  pages={1--8},
  year={2023},
  organization={IEEE},
  address={Budva, Montenegro},
  doi={10.1109/MECO58584.2023.10154913}
}

@article{andreasyan2026riscv,
  title={RISC-V Functional Safety for Autonomous Automotive Systems: An Analytical Framework and Research Roadmap for ML-Assisted Certification},
  author={Andreasyan, Nick and Struve, Mikhail and Popov, Alexey and Nikolaev, Maksim and Vashkelis, Vadim},
  journal={arXiv preprint arXiv:2604.17391},
  year={2026},
  doi={10.48550/arXiv.2604.17391}
}

@inproceedings{chen2024sac,
  title={SAC-KG: Exploiting large language models as skilled automatic constructors for domain knowledge graph},
  author={Chen, Hanzhu and Shen, Xu and Lv, Qitan and Wang, Jie and Ni, Xiaoqi and Ye, Jieping},
  booktitle={Proceedings of the 62nd Annual Meeting of the Association for Computational Linguistics (Volume 1: Long Papers)},
  pages={4345--4360},
  year={2024}
}

@inproceedings{papaluca2024zero,
  title={Zero-and few-shots knowledge graph triplet extraction with large language models},
  author={Papaluca, Andrea and Krefl, Daniel and M{\'e}ndez, Sergio Rodr{\'\i}guez and Lensky, Artem and Suominen, Hanna},
  booktitle={Proceedings of the 1st workshop on knowledge graphs and large language models (kaLLM 2024)},
  pages={12--23},
  year={2024}
}

@inproceedings{anuyah2025automated,
  title={Automated knowledge graph construction using large language models and sentence complexity modelling},
  author={Anuyah, Sydney and Kaushik, Mehedi Mahmud and Dwarampudi, Sri Rama Krishna Reddy and Shiradkar, Rakesh and Durresi, Arjan and Chakraborty, Sunandan},
  booktitle={Proceedings of the 2025 Conference on Empirical Methods in Natural Language Processing},
  pages={15526--15550},
  year={2025}
}

@inproceedings{niu2025tree,
  title={Tree-KG: An expandable knowledge graph construction framework for knowledge-intensive domains},
  author={Niu, Songjie and Yang, Kaisen and Zhao, Rui and Liu, Yichao and Li, Zonglin and Wang, Hongning and Chen, Wenguang},
  booktitle={Proceedings of the 63rd Annual Meeting of the Association for Computational Linguistics (Volume 1: Long Papers)},
  pages={18516--18529},
  year={2025}
}

@inproceedings{liu2025aligning,
  title={Aligning vision to language: Annotation-free multimodal knowledge graph construction for enhanced llms reasoning},
  author={Liu, Junming and Meng, Siyuan and Gao, Yanting and Mao, Song and Cai, Pinlong and Yan, Guohang and Chen, Yirong and Bian, Zilin and Wang, Ding and Shi, Botian},
  booktitle={Proceedings of the IEEE/CVF International Conference on Computer Vision},
  pages={981--992},
  year={2025}
}

@inproceedings{bu2025query,
  title={Query-driven multimodal GraphRAG: Dynamic local knowledge graph construction for online reasoning},
  author={Bu, Chenyang and Chang, Guojie and Chen, Zihao and Dang, CunYuan and Wu, Zhize and He, Yi and Wu, Xindong},
  booktitle={Findings of the Association for Computational Linguistics: ACL 2025},
  pages={21360--21380},
  year={2025}
}

@article{almahri2026enhancing,
  title={Enhancing supply chain visibility with knowledge graphs and large language models},
  author={AlMahri, Sara and Xu, Liming and Brintrup, Alexandra},
  journal={International Journal of Production Research},
  volume={64},
  number={6},
  pages={2178--2209},
  year={2026},
  publisher={Taylor \& Francis}
}

@inproceedings{han2024critical,
  title={Critical Path Identification in Supply Chain Knowledge Graphs with Large Language Models},
  author={Han, Yaomengxi and Ding, Zifeng and Liu, Yushan and He, Bailan and Tresp, Volker},
  booktitle={European Semantic Web Conference},
  pages={223--227},
  year={2024},
  organization={Springer}
}

@article{shahsavari2024event,
  title={Event identification for supply chain risk management through news analysis by using large language models},
  author={Shahsavari, Maryam and Hussain, Omar Khadeer and Saberi, Morteza and Sharma, Pankaj},
  journal={The Review of Socionetwork Strategies},
  volume={18},
  number={2},
  pages={255--278},
  year={2024},
  publisher={Springer}
}

@inproceedings{cheng2024shield,
  title={Shield: Llm-driven schema induction for predictive analytics in ev battery supply chain disruptions},
  author={Cheng, Zhi-Qi and Dong, Yifei and Shi, Aike and Liu, Wei and Hu, Yuzhi and O’Connor, Jason and Hauptmann, Alexander G and Whitefoot, Kate},
  booktitle={Proceedings of the 2024 Conference on Empirical Methods in Natural Language Processing: Industry Track},
  pages={303--333},
  year={2024}
}

@article{li2023large,
  title={Large language models for supply chain optimization},
  author={Li, Beibin and Mellou, Konstantina and Zhang, Bo and Pathuri, Jeevan and Menache, Ishai},
  journal={arXiv preprint arXiv:2307.03875},
  year={2023}
}

@article{simchi2025large,
  title={Large language models for supply chain decisions},
  author={Simchi-Levi, David and Mellou, Konstantina and Menache, Ishai and Pathuri, Jeevan},
  journal={arXiv preprint arXiv:2507.21502},
  year={2025}
}

@article{song2026large,
  title={Large language models in supply chain management: a systematic literature review and application framework},
  author={Song, Zhe and Xie, Ying and Yang, Lichao and Zhao, Yifan},
  journal={International Journal of Production Research},
  pages={1--41},
  year={2026},
  publisher={Taylor \& Francis}
}

@article{singh2026generative,
  title={Generative artificial intelligence (GenAI) in procurement and supply chain management: applications, opportunities and challenges},
  author={Singh, Vinay and Hughes, Laurie and Albashrawi, Mousa Ahmed and Jeon, Il and Dwivedi, Yogesh K},
  journal={Journal of Systems and Information Technology},
  volume={28},
  number={1},
  pages={123--144},
  year={2026},
  publisher={Emerald Publishing Limited}
}

@article{almahri2026automating,
  title={Automating Supply Chain Disruption Monitoring via an Agentic AI Approach},
  author={AlMahri, Sara and Xu, Liming and Brintrup, Alexandra},
  journal={arXiv preprint arXiv:2601.09680},
  year={2026}
}

@inproceedings{wang2024docllm,
  title={Docllm: A layout-aware generative language model for multimodal document understanding},
  author={Wang, Dongsheng and Raman, Natraj and Sibue, Mathieu and Ma, Zhiqiang and Babkin, Petr and Kaur, Simerjot and Pei, Yulong and Nourbakhsh, Armineh and Liu, Xiaomo},
  booktitle={Proceedings of the 62nd Annual Meeting of the Association for Computational Linguistics (Volume 1: Long Papers)},
  pages={8529--8548},
  year={2024}
}

@article{liu2025structured,
  title={Structured attention matters to multimodal llms in document understanding},
  author={Liu, Chang and Chen, Hongkai and Cai, Yujun and Wu, Hang and Ye, Qingwen and Yang, Ming-Hsuan and Wang, Yiwei},
  journal={arXiv preprint arXiv:2506.21600},
  year={2025}
}

@inproceedings{zhang2024extract,
  title={Extract, define, canonicalize: An llm-based framework for knowledge graph construction},
  author={Zhang, Bowen and Soh, Harold},
  booktitle={Proceedings of the 2024 conference on empirical methods in natural language processing},
  pages={9820--9836},
  year={2024}
}

@article{bai2025autoschemakg,
  title={Autoschemakg: Autonomous knowledge graph construction through dynamic schema induction from web-scale corpora},
  author={Bai, Jiaxin and Fan, Wei and Hu, Qi and Zong, Qing and Li, Chunyang and Tsang, Hong Ting and Luo, Hongyu and Yim, Yauwai and Huang, Haoyu and Zhou, Xiao and others},
  journal={arXiv preprint arXiv:2505.23628},
  year={2025}
}

@article{sun2025lkd,
  title={LKD-KGC: Domain-specific KG construction via LLM-driven knowledge dependency parsing},
  author={Sun, Jiaqi and Qian, Shiyou and Han, Zhangchi and Li, Wei and Qian, Zelin and Yang, Dingyu and Cao, Jian and Xue, Guangtao},
  journal={arXiv preprint arXiv:2505.24163},
  year={2025}
}

@article{mo2025kggen,
  title={Kggen: Extracting knowledge graphs from plain text with language models},
  author={Mo, Belinda and Yu, Kyssen and Kazdan, Joshua and Cabezas, Joan and Mpala, Proud and Yu, Lisa and Cundy, Chris and Kanatsoulis, Charilaos and Koyejo, Sanmi},
  journal={arXiv preprint arXiv:2502.09956},
  year={2025}
}

@article{li2025integrating,
  title={Integrating graph retrieval-augmented generation with large language models for supplier discovery},
  author={Li, Yunqing and Ko, Hyunwoong and Ameri, Farhad},
  journal={Journal of Computing and Information Science in Engineering},
  volume={25},
  number={2},
  pages={021010},
  year={2025},
  publisher={American Society of Mechanical Engineers}
}

@article{zheng2025enhancing,
  title={Enhancing supply chain visibility with generative AI: an exploratory case study on relationship prediction in knowledge graphs},
  author={Zheng, Ge and Brintrup, Alexandra},
  journal={International Journal of Production Research},
  pages={1--23},
  year={2025},
  publisher={Taylor \& Francis}
}

@inproceedings{lairgi2024itext2kg,
  title={itext2kg: Incremental knowledge graphs construction using large language models},
  author={Lairgi, Yassir and Moncla, Ludovic and Cazabet, R{\'e}my and Benabdeslem, Khalid and Cl{\'e}au, Pierre},
  booktitle={International Conference on Web Information Systems Engineering},
  pages={214--229},
  year={2024},
  organization={Springer}
}

@inproceedings{petrovic2025llm,
  title={Llm-based iterative approach to metamodeling in automotive},
  author={Petrovic, Nenad and Pan, Fengjunjie and Zolfaghari, Vahid and Knoll, Alois},
  booktitle={2025 2nd International Generative AI and Computational Language Modelling Conference (GACLM)},
  pages={266--271},
  year={2025},
  organization={IEEE}
}

@article{petrovic2025llmEC,
  title={LLM-Empowered Event-Chain Driven Code Generation for ADAS in SDV systems},
  author={Petrovic, Nenad and Kroth, Norbert and Torschmied, Axel and Song, Yinglei and Pan, Fengjunjie and Zolfaghari, Vahid and Purschke, Nils and Kirchner, Sven and Wu, Chengdong and Schamschurko, Andre and Zhang, Yi and Knoll, Alois},
  journal={arXiv preprint arXiv:2511.21877},
  year={2025}
}

@article{petrovic2025genai,
  title={GenAI for Automotive Software Development: From Requirements to Wheels},
  author={Petrovic, N. and Pan, F. and Zolfaghari, V. and Lebioda, K. and Schamschurko, A. and Knoll, A.},
  journal={arXiv preprint arXiv:2507.18223},
  year={2025}
}
\end{document}